# Micro-Actuators and Implementation

Pavan B Garadi[1] , Sai Bhargav M[2], Harishesh J K[3], Mamatha A S[4]

Dept. of ECE, RNS Institute of Technology, Bengaluru

[1]pavan.garadi96@gmail.com, [2]saibhargav1508@gmail.com, [3] harishesh97@gmail.com , [4]mamathatanya@gmail.com

*Abstract*— **Miniaturization of devices inculcates the need for small-sized actuators. Actuators in the size of a few centimeters are not uncommon but miniature devices need ones that are less than a few centimeters in dimension. Market for such actuators is rather small and information pertaining to their implementation is limited. This paper talks about various actuators and their actuation mechanism for the design of small-scale electronic devices. Not only are the small-sized actuators used for designing miniature devices, but also used for precise movements in the range of a few millimeters. We have included a procedure-wise description on how to implement these actuators. An in-depth analysis of their mechanical, electrical and chemical characteristics is elaborated in this paper.**

*Keywords—Miniaturization, Micro Actuators, Actuation mechanism, Drive circuits.*

## I. INTRODUCTION

An actuator is a device that converts energy into action/motion. Electronic actuators take in electrical energy and induce physical motion. Actuators are divided through their shape and style based on the device requirement [1].

They cause three types of motion- linear, rotary and a combination of the two. Linear actuator [2] creates movement along a straight axis. The shaft is restricted to movement only along one axis which results in linear movement in a straight line. A rotary actuator produces rotational motion, resulting in torque. The principle involved is Fleming's left hand rule which states that "a current carrying conductor placed in a magnetic field experiences a mechanical force". The third type of actuation mechanism is similar to linear actuation but the only difference is that the resulting actuation is due to the translation of circular motion, on an inclined plane, into linear motion [3].

Micro technologies are a crucial key to system performance, such as in automotive technologies. When devices are miniaturized, concepts concerning physical force and reactions are different from the macro world. Regarding application of micro actuators, the design method and the machining technologies need improvement in order to increase the actuator performance due to the large differences in theoretical design and properties of the prototype.

Selecting the right linear actuator for particular application involves taking into account certain factors such as necessary speed, stroke, actuation length, load and so on [4].

Amongst different sizes, technologies, actuation mechanisms and qualities, one can choose from hundreds of options that are available in the market. The trick is to narrow down to the actuator that delivers the best of results.

Fortunately, performing a decisive study to choose the right actuator is not a tough task. The design requirements reduce the set of possible actuator solutions. Speed is an important factor to be considered, the reason being faster movement leads to premature wear and results in undesired vibrations and other discomforts. It is essential that the actuator must be designed appropriately to handle the desired load. There are several factors to be considered whilst sizing for load capacity: the radial load capacity of the guide bearings, the moment capacity of the support carriage, and the axial load capacity of the support bearings and the ball screw [5]. Repeatability and durability are some factors that weigh in while choosing an actuator. Some of the most common fields were miniature actuators are used are camera technology, surgical equipment, precision printing and robotics and even in Nano satellites.

## II. ACTUATORS AND IMPLEMENTATION

There are various kinds of actuators such as: Micro stepper motors, Micro solenoids, Flexinol (muscle wire and SMA), DC motor, Micro vibrator, Piezo actuator, EAP and DEAs. Such micro actuators have been developed by the combination of micro managing and conventional technologies. The following Fig.1 depicts the block diagram of a simple Actuator system (Actuator interfacing system).

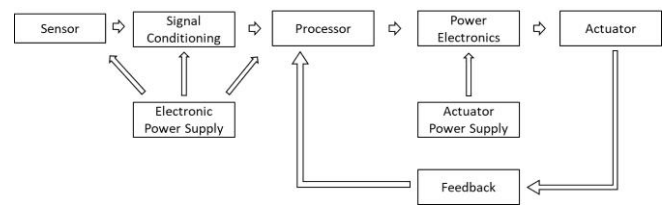

Figure 1: Block diagram of a simple actuator system

### A. Micro Stepper Motors

Stepper motors are basically D C motors that operate or move in steps [6]. Core principle of operation involves giving pulses for short durations unlike D C motors which require a constant D C source. Stepper motors are classified into 3 types [7] based on the type of rotor used – [i] Permanent magnet type: where the rotor is composed of a permanent magnet and working principle is simple attraction and repulsion between rotor magnets and the stator electromagnets. [ii] Variable Reluctance Type: has a plain iron rotor and operates on minimum reluctance. [iii] Hybrid Synchronous type: has a rotor mechanism which is obtained by combining Permanent magnet and variable reluctance techniques. Whatever be the type of winding, the excitation logic remains the same and can be applied to all.

**Properties**: There are two classes of stepper motors based on polarity. In other words polarity can be distinguished based on the windings. Unipolar stepper motors are those which have one winding with a center tap for each phase. Whereas every phase in a bipolar stepper motor has one winding [8].

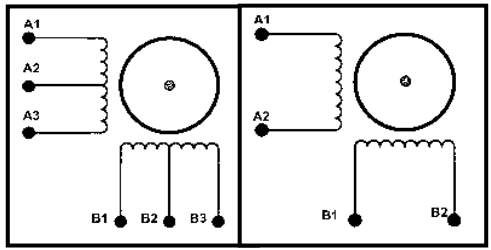

Figure 2: Unipolar and Bipolar Stepper motor coil windings and terminals

The most common type is the two phase stepper motor of either 3 wires or 6 wires. Within the stepper motor the centre taps may or may not be internally shorted, thereby making the 6 wire motor into a 5 wire 2 phase unipolar stepper motor.

**Implementation with diagrams**: Most miniature stepper motors are expected to have a voltage rating of about 3 to 6V maximum and draws current ranging from 10mA to a maximum of 50mA. The torque of a stepper is directly proportional to its current and voltage rating.

$$\theta r = \frac{360}{\text{No. of rotor pole teeth}} \quad \text{.........Eq. (1)}$$

$$\theta s = \frac{360}{\text{No. of stator pole teeth}} \quad \text{.........Eq. (2)}$$

As stated earlier, a stepper motor needs a sequence of pulses to operate. The pulse rate determines the speed of rotation as in Eq.1 and 2. The rotor advances in steps and each step is brought about by one single pulse. For example if a stepper motor has a step angle of 2 degrees, i.e. each pulse makes the rotor rotate by an angle of 2 degrees, then the motor needs a total of 180 pulses to make one complete rotation (360 degrees). Step angle varies depending on the type of stepper and type of internal winding. Therefore the pulses to be given have particular sequences depending upon the winding. Referring to Fig.2 the unipolar, 2-phase stepper motor coil excitation is as shown in the table below. The centre taps are shorted together and left unconnected. The A1, A3, B1 and B3 coils are excited as explained in table 1.

Table 1. Coil excitation table for a unipolar stepper motor.

| Step | Phase | | | |
|---|---|---|---|---|
| | A1 | B1 | A3 | B3 |
| 1 | 1 | 0 | 0 | 0 |
| 2 | 1 | 1 | 0 | 0 |
| 3 | 0 | 1 | 0 | 0 |
| 4 | 0 | 1 | 1 | 0 |
| 5 | 0 | 1 | 1 | 0 |
| 6 | 0 | 0 | 1 | 1 |
| 7 | 0 | 0 | 0 | 1 |
| 8 | 1 | 0 | 0 | 1 |

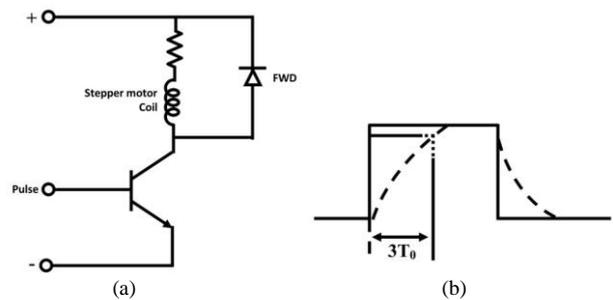

Figure 3: (a) Transistorized drive circuit for a unipolar stepper motor. (b) Timing diagram of actual and expected current in the conductor.

A rectangular pulse is applied to energize the coil by driving the base of the transistor. A freewheeling diode (FWD) is connected in parallel with the RL circuit to avoid back emf which arises during the transition from the positive to negative phase. The voltage is developed across the inductor gradually, with time constant $T_o$. The circuit is designed such that the pulse width is at least 6 to 8 times the time constant in order to leave the switching action unaffected.

**Applications:** Major advantage of a stepper motor is that the turn angle can be controlled as per requirement. The step increment can be converted from radial movement into liner increment by the use of a simple 90 degree gear system. Therefore a stepper motor can be used as a linear actuator as well. There are miniature stepper motors available in the market if the size of a few cms. Stepper motors can be used in the field of robotics where precision radial or liner movement is required. The precision movement is given by controlling the steps of the motor.

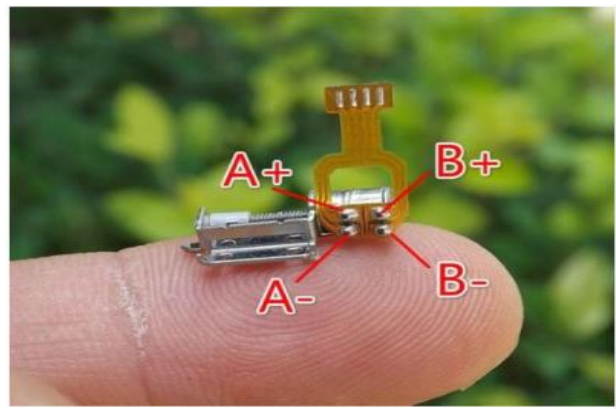

Figure 4: An example of a micro stepper motor available on an online forum

*B. Micro DC Motors*

DC motor is a machine that converts DC power into mechanical power. The basic operation is based on the principle that when a current carrying conductor is placed in a magnetic field the conductor experiences a mechanical force. The direction of the force is given by Fleming's left hand rule and magnitude is given by Eq.3 as,

$$F = B \times I \times L \text{ Newtons} \quad \text{............Eq. (3)}$$

There are three types of DC motors characterized by the connections of field winding in relation to the armature:

[i] Shunt Wound Motor: The field winding is connected in parallel with the armature. The current through the shunt field winding is not the same as an armature current.

[ii] Compound Wound Motor: This type has 2 field windings, one in parallel and another in series with the armature. The compound machines are always designed so that the flux produced by field winding is larger than the flux produced by the series winding.

[iii] Series Wound Motor: The field winding is connected in series with the armature. Series armature current runs through the series field winding and must be designed with fewer turns.

**Properties**: all DC motors have two terminals across which a constant DC voltage must be applied, and the motor spins in a particular direction. If the polarity of the applied DC voltage is reversed, then the DC motor starts spinning in the opposite direction. The speed with which a DC motor rotates depends upon the voltage applied and the load.

The most commonly used DC motors for small designs are the Brushless DC motors. In brushless DC motors, commutators are not used. Coil windings and direction of current in them determine the movement of rotor. Miniature Brushless motors are not uncommon in the field of drones and micro or nano satellites. Brushless motors are far easier to reduce size wise compared to other DC motors [9].

$$\begin{bmatrix} Va \\ Vb \\ Vc \end{bmatrix} = Rs \begin{bmatrix} ia \\ ib \\ ic \end{bmatrix} + L \frac{d}{dt} \begin{bmatrix} ia \\ ib \\ ic \end{bmatrix} + \begin{bmatrix} ea \\ eb \\ ec \end{bmatrix} \quad \text{...........Eq. (4)}$$

Where L – three – phase stator inductance, Va, Vb, Vc are voltages and ea, eb, ec are emfs in Eq. 4. As stated earlier, brushless DC motors rely on coil-rotor configuration and coil excitation to operate. The coils shall be excited sequentially. Internally, the coils are arranged in a star configuration with three pairs of coil endings as depicted in the below. The timing diagram depicted below in figure 5(c) shows the overlapping coil pulses.

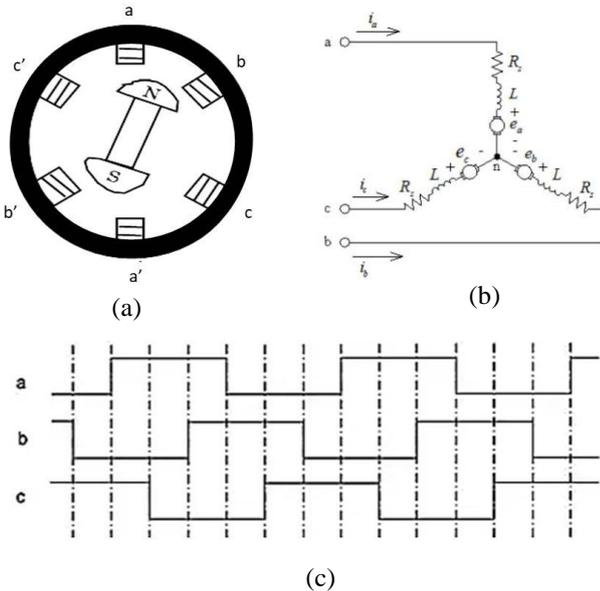

Figure 5: (a) Internal coil winding structure of a BLDC motor. (b) Coil Current distribution within the BLDC motor (c) Timing diagram of coil excitation of the BLDC motor

**Drive Circuit:** Due to the complex windings within the motor, the excitation logic is rather complex. Therefore the recommended option for a drive circuit is to implement a BLDC motor driver. The advantage of implementing a BLDC motor driver is that the task of designing a driver circuit is made effortless. The driver has control logic of its own due to a microcontroller embedded within the driver board. A good example for a BLDC motor driver is the STSPIN32F0/STSPIN32F0A (depicted in the figure 6 below).

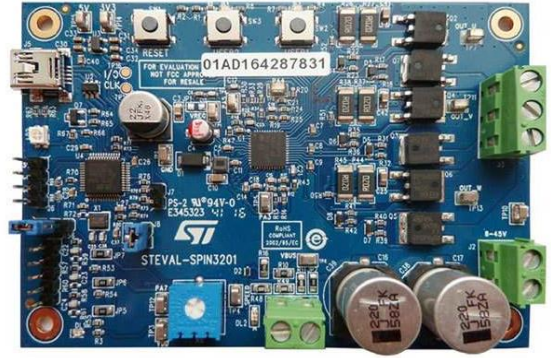

Figure 6: STSPIN32F0 BLDC motor driver

This STS motor driver implements a 32-bit ARM Cortex microcontroller and a typical 3-phase, half bridge gate driver and all these are embedded into a small package having dimensions of 7mm x 7mm. The small size of the driver presents the designer with an undue advantage of a miniature design possibility [10].

**Applications:** Powerful BLDC motors are some of the most common miniature motors that are available in the market. The main use of such small BLDC motors is in the field of drones. Due to the small size and minimum power requirements, these small BLDC motors are used as propeller motors in almost all drones. These DC motors have also found a use in the field of satellite technology. Micro satellites and nano satellites are so small that they are dependent entirely on miniature technology and hence suitable for the utilization of small DC motors. Satellites can put to use, these DC motors, for purposes like attitude and orbital control and for cameras on-board the satellites.

*C. Electro-Active Polymers (EAP)*

This is an electrostatic actuator for the laser speckle reducers. They (EAPs) are also known as 'artificial muscles' and are known to undergo large deformation while sustaining large forces. On a comparative note, EAP actuators have an input to deformation ratio which has a value that is much larger than the deformation ratio of piezo actuators. In comparison with other actuators, EAPs have certain advantages such as large deformation (strain >380%), high energy efficiency, large operating temperature range, and produces less noise and vibration.

$$s_2 = -\epsilon_r \epsilon_0 \frac{V^2}{Y\, t^2} \quad \text{...................Eq. (5)}$$

$$s_{planar} = \frac{1}{\sqrt{1 - \epsilon_r \epsilon_0 \frac{V^2}{Y t^2}}} - 1 \quad \text{............Eq. (6)}$$

Eq.5, 6 represent the thickness of the polymer, Y represents the modulus of elasticity, and the other two are constants.

**Working**: Electro-active polymers are driven by the principle called dielectric electro-active polymer principle. The main aspect of actuation depends on the electric field. Two electrodes sandwich a polymer and when a voltage is applied between the two electrodes, an electromechanical strain is introduced the existing electrostatic forces (Maxwell's stress).

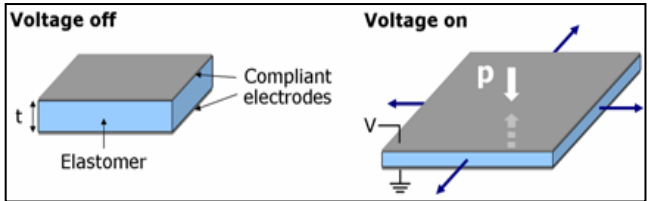

Figure 7: Diagrammatic representation of the actuation mechanism of EAPs.

The figure 7above represents the actuation mechanism of the EAP actuators. As stated earlier, application of a potential or voltage across two ends of the actuator (or to the two electrodes binding the polymer) results in the radial expansion of the polymer in the direction perpendicular to the application of voltage.

**Driving requirements**: EAPS are of different kinds, for example, electrostrictive, electrostatic, piezoelectric, and ferroelectric and so on. EAP material can result in displacement while operating under a DC voltage[11]. However electronic EAPs require high voltage fields greater than 10V/µm, whereas ionic EAPs can be activated by voltages as low as 1 to 2 V.

**Applications**: In comparison with other actuators, EAPs have certain advantages such as large deformation (strain >380%), high energy efficiency, large operating temperature range, and produces less noise and vibration. Some popular applications of EAPs miniature pumps and fluid valves, spring roll actuators in robotics, heel-strike generators and so on.

### D. Micro Solenoids

Solenoids are widely used among electromagnetic actuators because they have a simple structure with control in both direction and magnitude. The long loop of wire wrapped around the metal core produces magnetic flux and generates linear movement of the plunge when current is passed through the coil. Linear solenoids are both unidirectional and bi-directional; they often have a spring which pulls the plunger back to the home position[12].

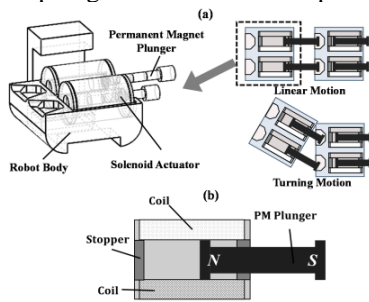

Figure 8: The design of a solenoid with a permanent magnet (PM) plunger: (a) the design concept of multi-segment robots with linear and turning motions using a pair of solenoids; (b) the cross-sectional view of the proposed solenoid actuator with PM plunger.

**Working**: The design requisites of a small-sized solenoid are determined using an optimization method. A prototype of the two-segmented robot equipped with two solenoids is used to realize linear and rotary motions and compare it with theoretical analysis.

The electromagnetic power produced by the solenoid is executed to structure a solenoid actuator with a lasting magnet plunger. A solenoid actuator is built for bi-directional movement in the scaled down robot. This solenoid actuator has a basic structure made out of curls and a lasting magnet plunger. The resultant power by the solenoid relies upon the length, air-hole, number of curl turns, connected current, and the sort of changeless magnet, and so forth. The condition of electromagnetic power activated by the solenoid is incited as the elements of these plan factors to build up a reasonable solenoid for fragmented robots. The magnetic field along the axis of a circular current loop of radius r and steady current can be expressed by the Biot-Savart's law as in Eq.7,

$$B_{sol} = (\mu_o \, i \, /4\pi) \int \ln(dl * r')/|r3| \quad \ldots\ldots\ldots \text{Eq. (7)}$$

The reaction normal for the actuator is estimated by changing the information voltage and by changing the recurrence of voltage input[13]. The graph illustrates the displacement achieved in the solenoid as a function of frequency. It is observed that there is maximum displacement of 1.8mm at 50Hz.

**Applications**: Solenoids play an important role in the field of linear actuation. These are some of the simplest mechanical and electrical designs that exist for liner actuators. The strength of actuation, speed, displacement and other factors can all be designed as per need. The design is solely based on the winding and core. Liner actuators can be used in a number of applications. For example, they can act as a valve for electrical circuitry. They can be used as terminals of contact in any circuit. The most futuristic application of the solenoid is in the Haptic touch, where the user gets a feedback from point of contact if there is a contact.

### E. Shape Memory Alloys

A shape-memory alloy (also referred as SMA) is an alloy that is deformed at low temperatures but returns to its shape (prior distortion) upon the application of heat. It is also called memory metal, memory alloy, smart metal, smart alloy, or muscle wire. The two most commonly used SMAs are aluminum-nickel, and nickel-titanium (NiTi) alloys but they can also be created by alloying compounds of zinc, copper, gold and iron in fixed percentages. Despite being iron-based or copper-based, SMAs such as Fe-Mn-Si, Cu-Zn-Al and Cu-Al-Ni, are commercially available and cheaper than Copper- based SMAs are preferred for their stability, practicability[14][15] and superior thermo-mechanic performance for most applications[16].

**Working**: The transition from the martensite to the austenite phase is reliant on temperature and stress but not time as phase changes are not in delusion. It is the reversible diffusion less transition between these two phases that results in special properties mainly used for actuation. Steel does not have shape-memory properties as martensite can be formed from austenite by rapidly cooling carbon-steel, this process is irreversible.

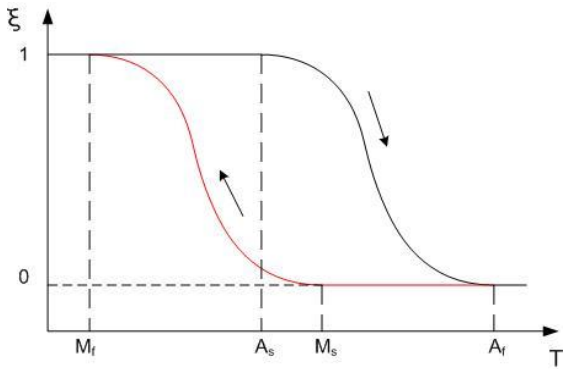

Figure 9: The graph represents the martensite fraction, ξ(T), when plotted against temperature, T.

In this figure, ξ(T) represents the martensite fraction. The difference between the heating and cooling transitions gives rise to hysteresis where some of the mechanical energy is lost during the transition. Shape-memory alloys are typically cast, using vacuum arc method or induction melting. These special techniques keep impurities to a minimum in the alloy and ensure that the metals are uniformly mixed. The cylindrical ingot is then hot rolled into larger sections and then drawn into thin wires. The ways in which the alloys are "developed" depend on the desired properties. The development procedure dictates the shape that the alloy will retain when heated. This occurs by heating the alloy so that the dislocations re-order themselves into stable structure, but not so hot that the material re-crystallizes. SMAs heated to temperatures between 400 °C to 500 °C for duration of 30 minutes, retaining their shape when hot, and then are cooled rapidly by blasting it with water or by cooling with air. The copper-based and NiTi-based shape-memory alloys are considered to be engineering materials and can be manufactured to almost any shape and size. The tensile strength of SMAs is lower than that of industrial steel, but some compositions have a higher strength than plastic or aluminum. The yield stress for Ni Ti can reach up to 500 MPa. The cost of manufacturing the metal is high and the processing requirements make it difficult and expensive to implement SMAs into practical designs. As a result, the materials are used in applications where their super elastic properties or shape-memory effects are exploited. The most common application is in linear actuation.

SMAs can induce high level of recoverable plastic strain reaching maximum recoverable strain these materials can hold without permanent damage up to 8% for some alloys. This is relative to steel with a maximum strain 0.5%. SMA actuators are usually actuated electrically, where an electric current leads to Joule heating. Deactivating the alloy is typically achieved by conductive heat transfer to the ambient environment. Therefore, SMA actuation is asymmetric, with a predominantly fast actuation time and a slow de-actuation time. Numerous methods have been proposed to reduce SMA deactivation time, including forced convection[17] and lagging the SMA with a conductive material in order to increase heat transfer rate to the environment. These actuators are made more feasible to include the use of a conductive "lagging". This method uses a thermal paste to rapidly transfer heat from the SMA by conduction due to which the heat is more readily transferred to the environment by convection as the outer radii are significantly greater than for the bare wire. This method reduces deactivation time significantly and gives a symmetric activation profile. Also, this method decreases required current to achieve a given actuation force[18].

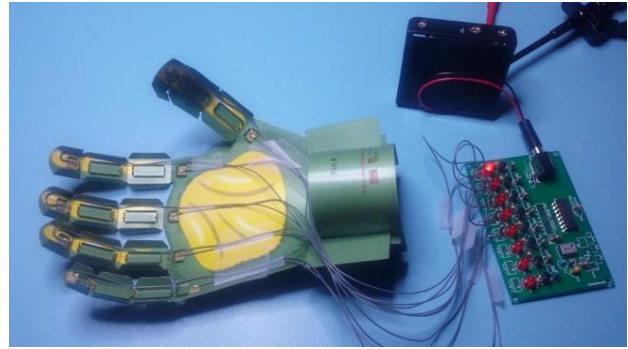

Figure 10: SMAs being used as muscle fibres in a robotic arm.

**Applications**: Some of the most profound use for SMAs can be in the field of medicine and Bio-Engineering. For example, Bones can be mended using SMAs and Clots can be removed within the arteries and veins. It can also be used in the field of dentistry where teeth can be fixed and placed using the dental wires made of SMA. The small size plays a major role in the field of surgical equipment.

*F. Micro Vibrators*

Precision Micro drives currently produces coin vibration motors, also known as shaft less or pancake vibrator motors, generally in Ø8mm - Ø12mm diameters for our Pico Vibe range. Pancake motors are compact and convenient to use.

**Working**: They integrate into many designs because they have no external moving parts, and can be affixed in place with a strong permanent self-adhesive mounting system. The coin or pancake vibrating motors is all Eccentric Rotating Mass (ERM) motors. Therefore they can be driven in the same manner as their pager motor counterparts. They have the same motor drive principles, including H-bridge circuitry for active braking. Brushed coin vibration motors are constructed from a flat PCB on which the 3-pole commutation circuit is laid out around an internal shaft in the centre[19].

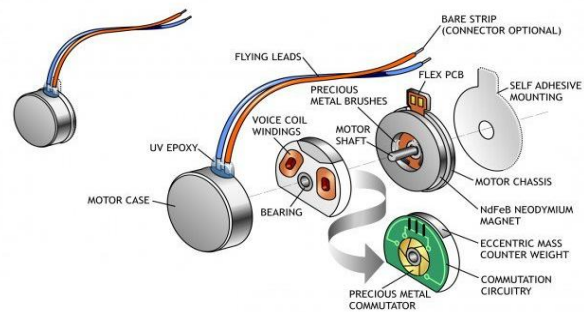

Figure 11: Internal structure of micro vibrator

The vibration motor rotor consists of two 'voice coils' and a small mass that is integrated into a flat plastic disc with a bearing in the middle, which sits on a shaft. Two brushes on the underside of the plastic disc make contact to the PCB commutation pads and provide power to the voice coils which generate a magnetic field. This field interacts with the flux generated by a disc magnet that is attached to the motor chassis. The commutation circuit alternates the direction of the field through the voice coils, and this interacts with the NS pole pairs that are built into the neodymium magnet. The graph in figure 11 illustrates that the motor increases its revolutions with increasing value of current as more magnetic flux is generated against the armature of the coils.

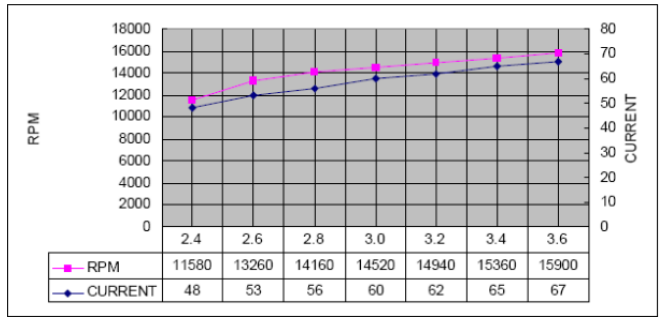

Figure 12: Rpm of the motor varies with change in current (µA).

**Driver circuit**: A recommended driver module or driver board would be the DRV2306[19]. The DRV is manufactured by Texas Instruments and is a reliable LRA (Linear Resonant Actuator) driver. The internal circuit is basically made up of a combination of H-Bridge and an amplifier. The input signal can be a PWM signal. The control logic is rather simple. A 100% duty cycle PWM to the input makes the micro vibrator motor rotate in one direction and a duty cycle of 0% makes the motor rotate in the opposite direction. A simplified block diagram of the DRV module interfaced with a Micro Vibrator is shown in the figure 13 below.

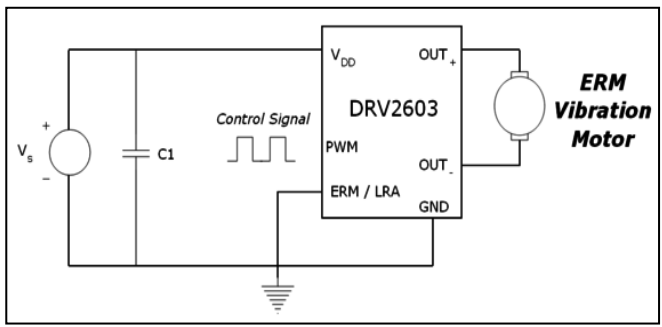

Figure 13: The DRV2603 Driver module interfaced with a Micro-Vibration Motor.

**Applications**: Due to their small size and enclosed vibration mechanism, coin vibrating motors are a popular choice for many different applications. They are great for haptic, particularly in handheld instruments where space can be at a premium. The following are some examples of where micro-vibrators can be used, Mobile Phones, RFID Scanners, Medical equipment, Industrial Precision Equipment and so on. Another notable application is in the field of haptic touch. The vibration produced by the motor acts as a feedback to the user upon contact. Micro vibrators are extensively used in mobile phones and such electronic devices

### III. CONCLUSION

When it comes to building miniature designs and miniature electronic devices, the most challenging task is to research about the right kind of actuators to be used in the design. Once the designer narrows down to a few actuators, taking a decisive step to choose the right actuator for the design proves to be a tough task. This paper enables the designer to look at the various micro actuators available in the market and helps in choosing the right one for the design. The work and research is made easier and the designer can focus on the design, electronics and software rather than focusing on what actuator is to be implemented. Not only does this paper describes the actuator, but also provides information on how to run and how to implement them in the desired manner. This paper provides the designer with an undue advantage over having to compromise for large scale designs. To summarize, the topics and concepts explained in this paper makes it easy to make a miniaturized electronic design for any device.

REFERENCES

[1] https://www.thomasnet.com/about/actuators-301168.html
[2] https://www.anaheimautomation.com/manuals/forms/linear-actuator-guide.php#sthash.dLxoWbWk.dpbs
[3] https://www.electronicdesign.com/electromechanical/linear-rotary-motion-combined-actuator/
[4] https://www.machinedesign.com/motion-control/how-select-right-linear-actuator
[5] https://www.mdpi.com/journal/actuators/
[6] https://learn.adafruit.com/all-about-stepper-motors/what-is-a-stepper-motor
[7] https://www.elprocus.com/stepper-motor-types-advantages-applications/
[8] http://www.robotiksistem.com/stepper_motor_types_properties.html
[9] https://ieeexplore.ieee.org/document/7368403
[10] https://www.st.com/en/motor-drivers/brushless-dc-motor-drivers.html
[11] https://iopscience.iop.org/article/10.1088/0964-1726/16/2/E01/pdf
[12] Lequesne, B. Fast-acting long-stroke bistable solenoid with moving permanent magnets. IEEE Trans. Ind. Appl. 1990, 26, 401–407.
[13] Proc. of The 1st IARP Workshop on Medical and Healthcare Robots, Ottawa, Canada, pp.789- 792, June 23 -24, (1988).
[14] Wilkes, K. E.; Liaw, P. K.; Wilkes, K. E. (2000). "The fatigue behavior of shape-memory alloys". JOM. 52 (10):45. Bibcode:2000JOM....52j..45W. doi:10.1007/s11837-000-0083-3.
[15] Cederström, J.; Van Humbeeck, J. (1995). "Relationship Between Shape Memory Material Properties and Applications" (PDF). Le Journal de Physique IV. 05: C2–335. doi:10.1051/jp4:1995251.
[16] Huang, W. (2002). "On the selection of shape memory alloys for actuators". Materials & Design. 23: 11-19. doi:10.1016/S0261-3069(01)00039-5.
[17] Lara-Quintanilla, A.; Hulskamp, A. W.; Bersee, H. E. (October 2013). "A high-rate shape memory alloy actuator for aerodynamic load control on wind turbines". Journal of Intelligent Material Systems and Structures. 24 (15): 1834 1845. doi:10.1177/1045389X13478271. Retrieved 12 November 2013.
[18] Huang, S; Leary, Martin; Attalla, Tamer; Probst, K; Subic, A (2012). "Optimisation of Ni–Ti shape memory alloy response time by transient heat transfer analysis". Materials & Design.
[19] https://www.precisionmicrodrives.com/content/ab-017-integrated-driver-circuits-for-vibration-motors/